\documentclass[a4paper,10pt]{article}
\usepackage[dvips]{graphicx}
\usepackage{amssymb,amsmath}
\numberwithin{equation}{section}

\begin{document}

 \newcommand{\be}[1]{\begin{equation}\label{#1}}
 \newcommand{\ee}{\end{equation}}
 \newcommand{\bea}{\begin{eqnarray}}
 \newcommand{\eea}{\end{eqnarray}}
 \def\disp{\displaystyle}

 \def\gsim{ \lower .75ex \hbox{$\sim$} \llap{\raise .27ex \hbox{$>$}} }
 \def\lsim{ \lower .75ex \hbox{$\sim$} \llap{\raise .27ex \hbox{$<$}} }



 \title{\Large \bf Metric-affine Myrzakulov gravity  and its generalizations}

\author{Ratbay Myrzakulov\footnote{Email: rmyrzakulov@gmail.com} \\ 
\textit{Ratbay Myrzakulov Eurasian International Center for Theoretical Physics, Astana 010009, Kazakhstan}}

\date{}
 \maketitle


 \renewcommand{\baselinestretch}{1.1}

 \begin{abstract}\vspace{1cm}
Since the discovery of cosmic acceleration, modified gravity theories play an important role in the
 modern cosmology. In particular, the well-known F(R)-theories reached great popularity motivated by the easier formalism and by the prospect to find a final theories of gravity for the dark scenarios.  In the present work, we study  some generalizations of F(R),  F(T) and F(Q)  gravity theories, where $R, T, Q$ are the Ricci, torsion and nonmetricity scalars.  At the beginning, we briefly review  the
 formalism of such  theories. Then, we will consider one of their generalizations, the so-called Myrzakulov F(R,T) gravity theory or the MG-I theory. The
 point-like Lagrangian is explicitly presented. Based on
 this Lagrangian, the field equations  of MG-I are found. For the specific model $F(R,T)=\mu R+\nu T,$  the
 corresponding exact solutions are derived. 
 Furthermore, we will consider the physical quantities associated 
 to such solutions and we will find how for some values of the parameters the expansion
 of our universe can be accelerated without introducing any dark component. Finally, some metric-affine Myrzakulov gravity theories with and without boundary term scalars are presented. 
 \end{abstract}

\tableofcontents

\section{Introduction}

Recent observational data imply -against any previous belief- that the current expansion of the universe
is accelerating~\cite{r1}. Since this discovery, the  so-called Dark Energy issue has probably become the most ambitious and tantalizing field of research because of its implications in fundamental physics. There exist several descriptions of the cosmic acceleration. Among them, the simplest one is the introduction of small positive 
Cosmological Constant in the framework of General Relativity (GR), the so-called $\Lambda$CDM Model, but it is well accepted the idea according to
which this is not the ultimate theory of gravity,
but an extremely good approximation valid in the present day
range of detection.
A generalization of this simple modification of GR consists in considering  modified gravitational theories~\cite{r1,r2}. In the last years the interest in modified gravity theories like $F(R)$ and $F(G)$-gravity as alternatives to the $\Lambda$CDM Model grew up.

Recently, a new
modified gravity theory, namely the $F(T)$-theory,
 has been proposed. This is a generalized version
 of the teleparallel gravity originally proposed by
 Einstein~\cite{r5}-\cite{MK3}. It also may describe  the current cosmic acceleration
 without invoking dark energy. Unlike the framework of GR, where the
Levi-Civita connection is used, in teleparallel gravity (TG) the used connection is the
Weitzenb\"ock'one. In principle, modification of gravity may contain a huge list of invariants and there is not any reason to restrict the
gravitational theory  to  GR, TG, $F(R)$ gravity and/or $F(T)$ gravity. Indeed, several
generalizations of these theories have been
proposed (see e.g. the quite recent review \cite{Ba}).
In this paper, we study some other generalizations of $F(R)$ and $F(T)$ gravity theories. At the beginning,  we briefly review the formalism of $F(R)$  gravity and $F(T)$ gravity in Friedmann-Robertson-Walker
 (FRW) universe.  The flat FRW space-time is described by the metric
 \begin{equation}\label{1.1}
 ds^2=-dt^2+a^2(t)(dx^2+dy^2+dz^2),
 \end{equation}
 where $a=a(t)$ is the scale factor. The orthonormal tetrad
 components $e_i(x^\mu)$ are related to the metric through
 \begin{equation}\label{1.2}
 g_{\mu\nu}=\eta_{ij}e_\mu^i e_\nu^j\,,
 \end{equation}
 where the Latin indices $i$, $j$ run over 0...3 for
 the tangent space of the manifold, while the Greek letters $\mu$,~$\nu$ are
 the coordinate indices on the manifold, also running over 0...3. 
 
$F(R)$ and $F(T)$ modified theories of gravity have been extensively
explored and the possibility to construct viable models in their frameworks has
been carefully analyzed in several papers (see \cite{Ba} for a recent review).  For such theories, the
physical motivations are principally
related to the possibility to reach a more realistic representation of
the gravitational fields near curvature singularities and to 
create some first order approximation for the quantum theory of
gravitational fields.  Recently, it has been registred a renaissance of
$F(R)$ and $F(T)$  gravity theories in the attempt to explain the late-time accelerated expansion of the
Universe \cite{Ba}.
In the modern cosmology, in order to construct (generalized) gravity theories, three quantities -- the curvature scalar,  the Gauss --Bonnet scalar and the torsion scalar -- are  usually used (about our notations see below): 
 \begin{eqnarray}
 R_s&=&g^{\mu\nu}R_{\mu\nu},\label{1.3}\\
 G_s&=& R^2 - 4R^{\mu\nu}R_{\mu\nu} + R^{\mu\nu\rho\sigma}R_{\mu\nu\rho\sigma},\label{1.4}\\
  T_s&=&{S_\rho}^{\mu\nu}\,{T^\rho}_{\mu\nu}.\label{1.5}
 \end{eqnarray}  
 In this paper, our aim  is to replace these quantities with the other three variables  
 in the form 
 \begin{eqnarray}
 R&=&u+g^{\mu\nu}R_{\mu\nu},\label{1.6}\\
 G&=& w+R^2 - 4R^{\mu\nu}R_{\mu\nu} + R^{\mu\nu\rho\sigma}R_{\mu\nu\rho\sigma},\label{1.7}\\
  T&=&v+{S_\rho}^{\mu\nu}\,{T^\rho}_{\mu\nu},\label{1.8}
 \end{eqnarray}
 where $u=u(x_i; g_{ij}, \dot{g_{ij}},\ddot{g_{ij}}, ... ; f_j)$, $v=v(x_i; g_{ij}, \dot{g_{ij}},\ddot{g_{ij}}, ... ; g_j)$ and  $w=w(x_i; g_{ij}, \dot{g_{ij}},\ddot{g_{ij}}, ... ; h_j)$ are some functions to be defined. As a result, we obtain some generalizations of the known modified gravity theories.
 With the FRW metric ansatz the three variables (\ref{1.3})-(\ref{1.5}) become
 \begin{eqnarray}
 R_s&=&6(\dot{H}+2H^2),\label{1.9}\\
 G_s&=& 24H^2(\dot{H}+H^2),\label{1.10}\\
  T_s&=&-6H^2,\label{1.11}
 \end{eqnarray}  
 where $H=(\ln a)_{t}$. In the contrast, in this paper we will use the following three variables 
 \begin{eqnarray}
 R&=&u+6(\dot{H}+2H^2),\label{1.12}\\
 G&=& w+24H^2(\dot{H}+H^2),\label{1.13}\\
  T&=&v-6H^2.\label{1.14}
 \end{eqnarray}
 
This paper is organized as follows. In Sec. 2, we
 briefly review the formalism of $F(R)$,  $F(T)$ and $F(G)$ gravity for FRW metric.
 In particular, the corresponding Lagrangians  are  explicitly
 presented. In Sec. 3, we consider $F(R,T)$ theory, where $R$ and $T$ will be generalized with respect to the usual notions of curvature scalar and  torsion scalar. Some reductions of $F(R,T)$ gravity are presented in Sec. 4.   In Sec. 5, the  specific model $F(R,T)=\mu R+\nu T$  is {\bf analized} and in Sec. 6 the exact power-law solution is found;  some  cosmological implications of the model will be here discussed.  The Bianchi type I version of $F(R,T)$ gravity is considered in Sec. 7. Sec. 8 is devoted to some generalizations of some modified gravity theories. The other examples of metric-affine Myrzakulov gravity theories were presented in Sec. 9. In sections 10, 11 and 12, we briefly mentioned our approach to the problem for the cosmological, 
spherically symmetric (black hole) solutions in MG including theories with the boundary term scalars. 
Final conclusions and remarks are provided in Sec. 13.

 \section{Preliminaries of $F(R)$, $F(G)$ and  $F(T)$ gravities}
 At the beginning, we present the basic equations of $F(R)$, $F(T)$ and $F(G)$ modified gravity theories. For simplicity we mainly work in  the FRW spacetime.

 \subsection{$F(R)$ gravity}
 The action of $F(R)$ theory is given by
  \begin{equation}\label{2.1}
 {\cal S}_{R}=\int d^4 x e[F(R)+L_m],
 \end{equation}
 where $R$ is the curvature scalar. We work with  the FRW metric \eqref{1.1}. In this case  $R$ assumes the form
 \begin{equation} \label{2.2}
R=R_s=6(\dot{H}+2H^2). 
\end{equation}
The action we rewrite as
 \begin{equation}\label{2.3}
 {\cal S}_R=\int dtL_R,
 \end{equation}
 where the   Lagrangian is given by
 \begin{equation}\label{2.4}
  L_R=a^3(F-RF_R)-
 6F_R a\dot{a}^2-6F_{RR}\dot{R}a^2\dot{a}-a^3L_m.
 \end{equation}
 The corresponding field equations of $F(R)$ gravity read
 \begin{eqnarray}
 6\dot{R}HF_{RR}-(R-6H^2)F_R+F&=&\rho,\label{2.5}\\
   -2\dot{R}^2F_{RRR}+[-4\dot{R}H-2\ddot{R}]F_{RR}+[-2H^2-4a^{-1} \ddot{a}+R]F_{R}-F &=&p,\label{2.6}\\
 \dot{\rho}+3H(\rho+p)&=&0.\label{2.7}
 \end{eqnarray}

 \subsection{$F(T)$ gravity}
In the modified teleparallel gravity, the gravitational action is
 \begin{equation}\label{2.8}
 {\cal S}_T=\int d^4 xe [F(T)+L_m],
 \end{equation}
 where $e={\rm det}\,(e_\mu^i)=\sqrt{-g}\,$, and for
 convenience we use the units $16\pi G=\hbar=c=1$ throughout.
 The torsion scalar $T$ is defined as
 \begin{equation}\label{2.9}
 T\equiv{S_\rho}^{\mu\nu}\,{T^\rho}_{\mu\nu}\,,
 \end{equation}
 where
 \begin{eqnarray}
 {T^\rho}_{\mu\nu} &\equiv &-e^\rho_i\left(\partial_\mu e^i_\nu
 -\partial_\nu e^i_\mu\right)\,,\label{2.10}\\
 {K^{\mu\nu}}_\rho &\equiv &-\frac{1}{2}\left({T^{\mu\nu}}_\rho
 -{T^{\nu\mu}}_\rho-{T_\rho}^{\mu\nu}\right)\,,\label{eq6}\label{2.11}\\
 {S_\rho}^{\mu\nu} &\equiv &\frac{1}{2}\left({K^{\mu\nu}}_\rho
 +\delta^\mu_\rho {T^{\theta\nu}}_\theta-
 \delta^\nu_\rho {T^{\theta\mu}}_\theta\right)\,.\label{2.12}
 \end{eqnarray}
 For a spatially flat FRW metric \eqref{1.1}, as a consequence of equations (\ref{2.9})
 and \eqref{1.1}, we have that the torsion scalar  assumes the form
 \begin{equation}\label{2.13}
 T=T_s=-6H^2.
 \end{equation}
 The action \eqref{2.8} can be written as
 \begin{equation}\label{2.14}
 {\cal S}_T=\int dt L_T,
 \end{equation}
where
 the point-like Lagrangian reads 
 \begin{equation}\label{2.15}
  L_T=a^3\left(F-F_T T\right)-
 6F_T a\dot{a}^2-a^3L_m.
 \end{equation}
  The equations of F(T) gravity look like
 \begin{eqnarray}
 12H^2 F_T+F&=&\rho,\label{2.16}\\
 48H^2 F_{TT}\dot{H}-F_T\left(12H^2+4\dot{H}\right)-F
 &=&p,\label{2.17}\\
 \dot{\rho}+3H(\rho+p)&=&0.\label{2.18}
 \end{eqnarray}

 \subsection{$F(G)$ gravity}
 The action of $F(G)$ theory is given by
 \begin{equation}\label{2.19}
 {\cal S}_{G}=\int d^4 x e[F(G)+L_m],
 \end{equation}
 where the Gauss -- Bonnet scalar $G$ for the FRW metric is
 \begin{equation} \label{2.20}
G=G_s=24H^2(\dot{H}+H^2). 
\end{equation}

\section{ A naive model of $F(R,T)$ gravity}

Our aim in this section is to present  a naive version   of $F(R,T)$ gravity. We assume that  the relevant action of $F(R,T)$ theory is given by \cite{MK1}
 \begin{equation}\label{3.1}
 {\cal S}_{37}=\int d^4 xe[F(R,T)+L_m],
 \end{equation}
 where $R=u+R_s$ and  $T=v+T_s$ are some dynamical geometrical variables to be defined, and $R_s$ and  $T_s$ are the usual curvature scalar and the torsion scalar for the FRW spacetime. It is the so-called M$_{37}$ - model \cite{MK1}. In this paper we will restrict ourselves to the simple case where for FRW spacetime $R$ and $T$ are given by
 \begin{eqnarray} 
R&=&u+6(\dot{H}+2H^2)=u+R_s,\label{3.2}\\
T&=&v-6H^2=v+T_s.\label{3.3}
\end{eqnarray}
  As we can see these two variables ($R,T$) are some analogies (generalizations) of the usual curvature scalar $(R_s$) and torsion scalar ($T_s$) and for obvious reasons  we will still continue to call them as  the "curvature" scalar" and the "torsion"  scalar. We note that, in general, $u=u(t, a,\dot{a}, \ddot{a},\dddot{a}, ...; f_i)$ and $v=v(t,a,\dot{a}, \ddot{a},\dddot{a}, ...; g_i)$ are some real functions, $H=(\ln a)_t$,  while $f_i$ and $g_i$ are some unknown functions related with the geometry of the spacetime.  Finally we can write the M$_{37}$ - model for the FRW spacetime as
\begin{eqnarray} 
 S_{37}&=&\int d^4 xe[F(R,T)+L_m],\label{3.4}\\
R&=&u+6(\dot{H}+2H^2),\label{3.5}\\
T&=&v-6H^2.\label{3.6}
\end{eqnarray}
In this paper we restrict ourselves to the case $u=u(a,\dot{a})$ and $v=v(a,\dot{a})$. The scale factor $a(t)$, the curvature scalar  $R$
 and the torsion scalar $T(t)$ are taken as independent
 dynamical variables. Then, after some algebra the  action \eqref{3.4}  becomes
 \begin{equation}\label{3.7}
 {\cal S}_{37}=\int dt L,
 \end{equation}
 where the point-like Lagrangian is given by
 \begin{equation}\label{3.8}
  L_{37}=a^3(F-TF_T-RF_R+vF_T+uF_R)-
 6(F_R+F_T) a\dot{a}^2-6(F_{RR}\dot{R}+F_{RT}\dot{T})a^2\dot{a}-a^3L_m.
 \end{equation}
 The corresponding equations of the M$_{37}$ - model assume the form \cite{MK1}
  \begin{eqnarray}
 D_2F_{RR}+D_1F_R+JF_{RT}+E_1F_T+KF&=&-2a^3\rho,\nonumber \,\,\\
   U+B_2F_{TT}+B_1F_{T}+C_2F_{RRT}+C_1F_{RTT}+C_0F_{RT}+MF  &=&6a^2p,\label{3.9}\\
 \dot{\rho}+3H(\rho+p)&=&0.\nonumber
 \end{eqnarray}
 Here
 \begin{eqnarray}
 D_2&=&-6\dot{R}a^2\dot{a},\label{3.10}\\
   D_1&=&-6a \dot{a}^2+a^3u_{\dot{a}}\dot{a}-a^3(u-R),\label{3.11}\\ 
  J&=&-6a^2 \dot{a}\dot{T},\label{3.12}\\ 
 E_1&=&-6a \dot{a}^2+a^3v_{\dot{a}}\dot{a}-a^3(v-T),\label{3.13}\\
 K&=&-a^3\label{3.14}
 \end{eqnarray}
 and
 \begin{eqnarray}
 U&=&A_3F_{RRR}+A_2F_{RR}+A_1F_{R},\nonumber\\
  A_3&=&-6\dot{R}^2a^2,\label{3.15}\\
 A_2&=&-6\ddot{R}a^2-12\dot{R}a\dot{a}+a^3\dot{R}u_{\dot{a}},\label{3.16}\\
   A_1&=&-6\dot{a}^2-12a \ddot{a}+3a^2\dot{a}u_{\dot{a}}+a^3\dot{u}_{\dot{a}}-3a^2(u-R)-a^3u_a,\label{3.17}\\
 B_2&=&-12\dot{T}a \dot{a}+a^3\dot{T}v_{\dot{a}},\label{3.18}\\ 
 B_1&=&-6\dot{a}^2-12a \ddot{a}+3a^2\dot{a}v_{\dot{a}}+a^3\dot{v}_{\dot{a}}-3a^2(v-T)-a^3v_a,\label{3.19}\\
 C_2&=&-12a^2\dot{R}\dot{T},\label{3.20}\\
 C_1&=&-6a^2\dot{T}^2,\label{3.21}\\
 C_0&=&-12\dot{R}a\dot{a}-12\dot{T}a\dot{a}-6a^2\ddot{T}+a^3\dot{R}v_{\dot{a}}+a^3\dot{T}u_{\dot{a}},\label{3.22}\\
 M&=&-3a^2.\label{3.23}
 \end{eqnarray} 
 We can rewrite the system \eqref{3.9} in terms of $H$ as
 \begin{eqnarray}
 DF_{RR}+D_1F_R+JF_{RT}+E_1F_T+KF&=&-2a^3\rho,\notag\\   U+B_2F_{TT}+B_1F_{T}+C_2F_{RRT}+C_1F_{RTT}+C_0F_{RT}+MF&=&6a^2p,\label{3.24}\\ 
 \dot{\rho}+3H(\rho+p)&=&0,\notag
 \end{eqnarray}
  where
 \begin{eqnarray}
 D_2&=&-6\dot{R}a^2\dot{a}=-6a^3H\dot{R},\label{3.25}\\
   D_1&=&-6a^3H^2+a^3u_{\dot{a}}\dot{a}+6a^3(\dot{H}+2H^2)=a^3u_{\dot{a}}\dot{a}+6a^3(\dot{H}+H^2),\label{3.26}\\ 
J&=&-6a^3 H\dot{T},\label{3.27}\\ 
E_1&=&-6a^3 H^2+a^3v_{\dot{a}}\dot{a}-6a^3H^2=-12a^3H^2+a^3v_{\dot{a}}\dot{a},\label{3.28}\\
 K&=&-a^3.\label{3.29}
 \end{eqnarray}
 and
 \begin{eqnarray}
  U&=&A_3F_{RRR}+A_2F_{RR}+A_1F_{R},\nonumber\\
 A_3&=&-6\dot{R}^2a^2,\label{3.30}\\
 A_2&=&-6\ddot{R}a^2-12\dot{R}a\dot{a}+a^3\dot{R}u_{\dot{a}}=-6\ddot{R}a^2-12\dot{R}a\dot{a}+a^3\dot{R}u_{\dot{a}},\label{3.31}\\
   A_1&=&-6\dot{a}^2-12a \ddot{a}+3a^2\dot{a}u_{\dot{a}}+a^3\dot{u}_{\dot{a}}-3a^2(u-R)-a^3u_a,\label{3.32}\\
 B_2&=&-12\dot{T}a \dot{a}+a^3\dot{T}v_{\dot{a}},\label{3.33}\\ 
 B_1&=&-6\dot{a}^2-12a \ddot{a}+3a^2\dot{a}v_{\dot{a}}+a^3\dot{v}_{\dot{a}}-3a^2(v-T)-a^3v_a,\label{3.34}\\
 C_2&=&-12a^2\dot{R}\dot{T},\label{3.35}\\
 C_1&=&-6a^2\dot{T}^2,\label{3.36}\\
 C_0&=&-12\dot{R}a\dot{a}-12\dot{T}a\dot{a}-6a^2\ddot{T}+a^3\dot{R}v_{\dot{a}}+a^3\dot{T}u_{\dot{a}},\label{3.37}\\
 M&=&-3a^2.\label{3.38}
 \end{eqnarray}

 \section{Reductions. Preliminary classification} Note that the system (\ref{3.9}) or \eqref{3.24} admits some  important reductions. Let us now present these particular cases.
 \subsection{Case: $F=R$} Now we consider the particular case  $F=R$. Thus, the   system \eqref{3.24} becomes 
 \begin{eqnarray}
 D_1+KR&=&-2a^3\rho,\nonumber\\
   A_1+MR &=&6a^2p,\label{4.1}\\
 \dot{\rho}+3H(\rho+p)&=&0\nonumber
 \end{eqnarray} or
 \begin{eqnarray}
 3H^2+0.5(u-\dot{a}u_{\dot{a}})&=&\rho,\nonumber\\   2\dot{H}+3H^2-0.5(\dot{a}u_{\dot{a}}+\frac{1}{3}a\dot{u}_{\dot{a}}-u)&=&-p,\label{4.2}\\
 \dot{\rho}+3H(\rho+p)&=&0.\nonumber
 \end{eqnarray}
Let us rewrite this system as
 \begin{eqnarray}
 3H^2&=&\rho+\rho_c,\nonumber\\
   2\dot{H}+3H^2&=&-(p+p_c),\label{4.3}\\
 \dot{\rho}+3H(\rho+p)&=&0,\nonumber
 \end{eqnarray}
 where
 \begin{eqnarray}
 \rho_c&=&-0.5(u-\dot{a}u_{\dot{a}}),\label{4.4}\\
   p_c&=&-0.5(\dot{a}u_{\dot{a}}+3^{-1}a\dot{u}_{\dot{a}}-u)\label{4.5}
 \end{eqnarray}
 are the corrections to the energy denisty and pressure. Note that  if $u=0$ we obtain the standard equations of  GR,
 
 \begin{eqnarray}
 3H^2&=&\rho,\nonumber \\
   2\dot{H}+3H^2&=&-p,\label{4.6}\\
 \dot{\rho}+3H(\rho+p)&=&0\nonumber
 \end{eqnarray}

 \subsection{Case: $F=T$} 
Let us now to consider $F=T$. Then the system \eqref{3.24} leads to
 \begin{eqnarray}
 E_1+KT&=&-2a^3\rho,\nonumber\\
   B_1+MT  &=&6a^2p,\label{4.7}\\
 \dot{\rho}-3H(\rho+p)&=&0,\nonumber
 \end{eqnarray}  or
 \begin{eqnarray}
 3H^2+0.5(v-\dot{a}v_{\dot{a}})&=&\rho,\nonumber \\
   2\dot{H}+3H^2-0.5(\dot{a}v_{\dot{a}}+\frac{1}{3}a\dot{v}_{\dot{a}}-v)&=&-p,\label{4.8}\\
 \dot{\rho}+3H(\rho+p)&=&0. \nonumber
 \end{eqnarray}
The above system can be rewritten as
 \begin{eqnarray}
 3H^2&=&\rho+\rho_c,\nonumber\\
   2\dot{H}+3H^2&=&-(p+p_c),\label{4.9}\\
 \dot{\rho}+3H(\rho+p)&=&0,\nonumber
 \end{eqnarray}
 where
 \begin{eqnarray}
 \rho_c&=&-0.5(v-\dot{a}v_{\dot{a}}),\label{4.10}\\
   p_c&=&-0.5(\dot{a}v_{\dot{a}}+3^{-1}a\dot{v}_{\dot{a}}-v)\label{4.11}
 \end{eqnarray}
 are the corrections to the energy density and pressure. Obviously, if $v=0$ we obtain the standard equations of GR \eqref{4.6}.

 \subsection{Case:  $F=F(T),\quad  u=v=0$} 

Let us take $F=F(T),\quad  u=v=0$. Then, the system \eqref{3.24} becomes
  \begin{eqnarray}
 E_1F_T+KF&=&-2a^3\rho,\label{4.12}\\
   B_2F_{TT}+B_1F_{T}+MF &=&6a^2p,\label{4.13}\\
 \dot{\rho}+3H(\rho+p)&=&0\label{4.14}
 \end{eqnarray}
 or
 \begin{eqnarray}
 -12a \dot{a}^2F_T-a^3F&=&-2a^3\rho,\label{4.15}\\
 -12\dot{T}a \dot{a}F_{TT}-(36\dot{a}^2+12a \ddot{a})F_{T}-3a^2F &=&6a^2p,\label{4.16}\\
 \dot{\rho}+3H(\rho+p)&=&0.\label{4.17}
 \end{eqnarray}
 This system can be rewritten as
 \begin{eqnarray}
 -2TF_T+F&=&2\rho,\label{4.18}\\
 -8\dot{H}TF_{TT}+2(T-2\dot{H})F_{T}-F &=&2p,\label{4.19}\\
 \dot{\rho}-3H(\rho+p)&=&0\label{4.20}
 \end{eqnarray}
 that is the same as \eqref{2.16}-\eqref{2.18} of $F(T)$ gravity.

\subsection{Case: $F=F(R), \quad u=v=0$} 
We get the second reduction if  we consider the case where $F=F(R), \quad u=v=0$. Then the system (\ref{3.9}) leads to  
 \begin{eqnarray}
 D_2F_{RR}+D_1F_R+KF&=&-2a^3\rho,\label{4.21}\\
   A_3F_{RRR}+A_2F_{RR}+A_1F_{R}+MF &=&6a^2p,\label{4.22}\\
 \dot{\rho}+3H(\rho+p)&=&0,\label{4.23}
 \end{eqnarray}
 where
 \begin{eqnarray}
 A_3&=&-6\dot{R}^2a^2,\label{4.24}\\
 A_2&=&-6\ddot{R}a^2-12\dot{R}a\dot{a},\label{4.25}\\
   A_1&=&-6\dot{a}^2-12a \ddot{a}+3a^2R,\label{4.26}\\
 D_2&=&-6\dot{R}a^2\dot{a},\label{4.27}\\
   D_1&=&-6a \dot{a}^2+a^3R,\label{4.28}\\
 K&=&-a^3.\label{4.29}
 \end{eqnarray}
  This system can be written as
  \begin{eqnarray}
 -6\dot{R}a^2\dot{a}F_{RR}+[-6a \dot{a}^2+a^3R]F_R-a ^3F&=&-2a^3\rho,\label{4.30}\\
   -6\dot{R}^2a^2F_{RRR}+[-12\dot{R}a \dot{a}-6\ddot{R}a^2]F_{RR}+[-6\dot{a}^2-12a \ddot{a}+3a^2R]F_{R}-3a^2F &=&6a^2p,\label{4.31}\\
 \dot{\rho}+3H(\rho+p)&=&0.\label{4.32}
 \end{eqnarray}
As a consequence,
 \begin{eqnarray}
 6\dot{R}HF_{RR}-(R-6H^2)F_R+F&=&2\rho,\label{4.33}\\
   -2\dot{R}^2F_{RRR}+[-4\dot{R}H-2\ddot{R}]F_{RR}+[-2H^2-4a^{-1} \ddot{a}+R]F_{R}-F &=&2p,\label{4.34}\\
 \dot{\rho}+3H(\rho+p)&=&0.\label{4.35}
 \end{eqnarray}
 This system corresponds to the one in equations (\ref{2.5})-(\ref{2.7}). We have shown that our model contents $F(R)$ and $F(T)$ gravity models as particular cases. In this sense it is the generalizations of these two known modified gravity theories.
 
 
 \subsection{The M$_{37A}$ - model}
 For the M$_{37A}$ - model we have $u\neq 0, \quad v=0$ so that
  \begin{eqnarray} 
 S_{37A}&=&\int d^4 xe[F(R,T)+L_m],\label{4.36}\\
R&=&u+6(\dot{H}+2H^2),\label{4.37}\\
T&=&-6H^2.\label{4.38}
\end{eqnarray}
 \subsection{The M$_{37B}$ - model}
 If we consider the case $u=0,\quad v\neq 0$, then we get  the M$_{37B}$ - model with
  \begin{eqnarray} 
 S_{37B}&=&\int d^4 xe[F(R,T)+L_m],\label{4.39}\\
R&=&6(\dot{H}+2H^2),\label{4.40}\\
T&=&v-6H^2.\label{4.41}
\end{eqnarray}

 \subsection{The M$_{37C}$ - model}
 Now we consider the case $v=\zeta(u)$. We get  the M$_{37C}$ - model with
 \begin{eqnarray} 
 S_{37B}&=&\int d^4 xe[F(R,T)+L_m],\label{4.42}\\
R&=&u+6(\dot{H}+2H^2),\label{4.43}\\
T&=&\zeta(u)-6H^2,\label{4.44}
\end{eqnarray}
where in general $\zeta$ is a function to be defined e.g. $\zeta=\zeta(t;a,\dot{a},\ddot{a}, \dddot{a}, ...; \varsigma;u)$ and $\varsigma$ is an unknown function.

 \subsection{The M$_{37D}$ - model}
 Now we consider the particular case of $u=\xi(v)$ and we get  the M$_{37D}$ - model with
 \begin{eqnarray}
 S_{37B}&=&\int d^4 xe[F(R,T)+L_m],\label{4.45}\\
R&=&\xi(v)+6(\dot{H}+2H^2),\label{4.46}\\
T&=&v-6H^2,\label{4.47}
\end{eqnarray}
where in general $\xi$ is a function to be defined e.g. $\xi=\xi(t;a,\dot{a},\ddot{a}, \dddot{a}, ...; \varsigma;v)$ and $\varsigma$ is an unknown function.

\subsection{The M$_{37E}$ - model}
 Finally  we consider the case  $u=v=0$ and we get  the M$_{37E}$ - model with
 \begin{eqnarray}
 S_{37E}&=&\int d^4 xe[F(R,T)+L_m],\label{4.48}\\
R&=&6(\dot{H}+2H^2),\label{4.49}\\
T&=&-6H^2.\label{4.50}
\end{eqnarray}
About this model we have some doubt related with the equation
\begin{equation}
 \dot{T}=-2(R+3T)\sqrt{-\frac{T}{6}}\label{4.51}
 \end{equation}
which follows from \eqref{4.49}-\eqref{4.50} by avoiding the variable $H$. This equation tell us that we have only one independent dynamical variable $R$ or $T$. It turns out that the model \eqref{4.48}-\eqref{4.50} is not of the type of $F(R,T)$ gravity, but is equivalent to $F(R)$ or $F(T)$ gravity only. This is why in this paper we introduced some new functions like $u, v$ and $w$ with the (temporally?)  unknown geometrical nature.

\subsection{The M$_{37F}$ - model}
 The M$_{37F}$ - model corresponds to the case
 \begin{equation}\label{4.52}
 R=0,\quad T\neq 0
 \end{equation}
 that is
 \begin{equation}
 u=-6(\dot{H}+2H^2)\label{4.53}
 \end{equation}
 As a consequence the M$_{37F}$ - model reads 
 \begin{eqnarray} \label{fe1}
 S_{37J}&=&\int d^4 xe[F(R,T)+L_m],\label{4.54}\\
R&=&0,\label{4.55}\\
T&=&v-6H^2.\label{4.56}
\end{eqnarray}
We can see that the M$_{37F}$ - model is in fact a generalization of $F(T)$ gravity.

\subsection{The M$_{37G}$ - model}
 We obtain the M$_{37G}$ - model by assuming
 \begin{equation}
 R\neq 0,\quad T= 0\label{4.57}
 \end{equation}
 that is
 \begin{equation}
 v=6H^2.\label{4.58}
 \end{equation}
 In this way we write the M$_{37G}$ - model as
 \begin{eqnarray} 
 S_{37J}&=&\int d^4 xe[F(R,T)+L_m],\label{4.59}\\
R&=&u+6(\dot{H}+2H^2),\label{4.60}\\
T&=&0.\label{4.61}
\end{eqnarray}
This model is in fact a generalization of $F(R)$ gravity.

\section{The particular model: $F(R,T)=\mu R+\nu T$}
The equations of $F(R,T)$ gravity are much more complicated with respect to the ones of GR even for FRW metric. For this reason let us consider the following simplest particular model
\begin{equation}\label{5.1}
 F(R,T)=\nu T+\mu R,
 \end{equation}
 where $\mu$ and $\nu$ are some real constants.  The equations system of $F(R,T)$ gravity becomes
 \begin{eqnarray}
 \mu D_1+\nu E_1+K(\nu T+\mu R)&=&-2a^3\rho,\label{5.2}\\
   \mu A_1+\nu B_1+M(\nu T+\mu R) &=&6a^2p,\label{5.3}\\
 \dot{\rho}+3H(\rho+p)&=&0,\label{5.4}
 \end{eqnarray}
 where
 \begin{eqnarray}
    D_1&=&-6a \dot{a}^2+a^3u_{\dot{a}}\dot{a}-a^3(u-R)=6a^2\ddot{a}+a^3\dot{a}u_{\dot{a}}=a^3(6\frac{\ddot{a}}{a}+\dot{a}u_{\dot{a}}),\label{5.5}\\ 
   E_1&=&-6a \dot{a}^2+a^3v_{\dot{a}}\dot{a}-a^3(v-T)=-12a \dot{a}^2+a^3\dot{a}v_{\dot{a}}=a^3(-12\frac{ \dot{a}^2}{a^2}+\dot{a}v_{\dot{a}}),\label{5.6}\\
 K&=&-a^3,\label{5.7}\\
   A_1&=&12\dot{a}^2+6a \ddot{a}+3a^2\dot{a}u_{\dot{a}}+a^3\dot{u}_{\dot{a}}-a^3u_a,\label{5.8}\\
  B_1&=&-24\dot{a}^2-12a \ddot{a}+3a^2\dot{a}v_{\dot{a}}+a^3\dot{v}_{\dot{a}}-a^3v_a,\label{5.9}\\
  M&=&-3a^2,\label{5.10}\\
 R&=&u+6\frac{\ddot{a}}{a}+6\frac{\dot{a}^2}{a^2}=u+6(\dot{H}+2H^2),\label{5.11}\\ 
 T&=&v-6\frac{\dot{a}^2}{a^2}=v-6H^2.\label{5.12}
 \end{eqnarray}
 We get\begin{eqnarray}
 -6(\mu+\nu)\frac{ \dot{a}^2}{a^2}+\mu \dot{a}u_{\dot{a}}+\nu \dot{a}v_{\dot{a}}-\mu u-\nu v&=&-2\rho,\label{5.13}\\
    -2(\mu+\nu)(\frac{ \dot{a}^2}{a^2}+2\frac{ \ddot{a}}{a})+\mu \dot{a}u_{\dot{a}}+\nu \dot{a}v_{\dot{a}}-\mu u-\nu v+\frac{\mu}{3}a(\dot{u}_{\dot{a}}-u_a)+\frac{\nu}{3}a(\dot{v}_{\dot{a}}-v_a)&=&2p,\label{5.14}\\
 \dot{\rho}+3H(\rho+p)&=&0.\label{5.15}
 \end{eqnarray}
 May rewrite it as
 \begin{eqnarray}
 3(\mu+\nu)\frac{ \dot{a}^2}{a^2}-0.5(\mu \dot{a}u_{\dot{a}}+\nu \dot{a}v_{\dot{a}}-\mu u-\nu v)&=&\rho,\label{5.16}\\
    (\mu+\nu)(\frac{ \dot{a}^2}{a^2}+2\frac{ \ddot{a}}{a})-0.5(\mu \dot{a}u_{\dot{a}}+\nu \dot{a}v_{\dot{a}}-\mu u-\nu v)-\frac{\mu}{6}a(\dot{u}_{\dot{a}}-u_a)-\frac{\nu}{6}a(\dot{v}_{\dot{a}}-v_a)&=&-p,\label{5.17}\\
 \dot{\rho}+3H(\rho+p)&=&0.\label{5.18}
 \end{eqnarray}
 or
 \begin{eqnarray}
 3(\mu+\nu)H^2-0.5(\mu \dot{a}u_{\dot{a}}+\nu \dot{a}v_{\dot{a}}-\mu u-\nu v)&=&\rho,\label{5.19}\\
    (\mu+\nu)(2\dot{H}+3H^2)-0.5(\mu \dot{a}u_{\dot{a}}+\nu \dot{a}v_{\dot{a}}-\mu u-\nu v)-\frac{\mu}{6}a(\dot{u}_{\dot{a}}-u_a)-\frac{\nu}{6}a(\dot{v}_{\dot{a}}-v_a)&=&-p,\label{5.20}\\
 \dot{\rho}-3H(\rho+p)&=&0.\label{5.21}
 \end{eqnarray}
 This system contents 2 equations and 5 unknown functions ($a,\rho, p, u, v$). 
 Note that  the EoS parameter is given by
 \begin{equation}\label{5.22}
 \omega= \frac{p}{\rho}=-1-\frac{2(\mu+\nu)\dot{H}-\frac{\mu}{6}a(\dot{u}_{\dot{a}}-u_a)-\frac{\nu}{6}a(\dot{v}_{\dot{a}}-v_a)}{3(\mu+\nu)H^2-0.5(\mu \dot{a}u_{\dot{a}}+\nu \dot{a}v_{\dot{a}}-\mu u-\nu v)}.
 \end{equation}
 Now we assume 
  \begin{equation}\label{5.23}
 u=\alpha a^n, \quad v=\beta a^m,
 \end{equation} 
 where $n, m, \alpha,\beta$ are some real constants so that we have 
  \begin{equation}\label{5.24}
 u=\alpha \left(\frac{v}{\beta}\right)^{\frac{n}{m}}, \quad v=\beta \left(\frac{u}{\alpha}\right)^{\frac{m}{n}},
 \end{equation} 
  Then, the previous system \eqref{5.16}-\eqref{5.18} leads to 
  \begin{eqnarray}
 3(\mu+\nu)\frac{ \dot{a}^2}{a^2}+0.5(\mu \alpha a^n+\nu \beta a^m)&=&\rho,\label{5.25}\\
    (\mu+\nu)(\frac{ \dot{a}^2}{a^2}+2\frac{ \ddot{a}}{a})+\frac{\mu\alpha (n+3)}{6}a^{n}+\frac{\nu\beta (m+3)}{6}a^m&=&-p,\label{5.26}\\
 \dot{\rho}+3H(\rho+p)&=&0\label{5.27}
 \end{eqnarray}
 or
 \begin{eqnarray}
 3(\mu+\nu)H^2+0.5(\mu \alpha a^n+\nu \beta a^m)&=&\rho,\label{5.28}\\
    (\mu+\nu)(2\dot{H}+3H^2)+\frac{\mu\alpha (n+3)}{6}a^{n}+\frac{\nu\beta (m+3)}{6}a^m&=&-p,\label{5.29}\\
 \dot{\rho}+3H(\rho+p)&=&0.\label{5.30}
 \end{eqnarray}

 
 \section{Cosmological  implications. Dark energy}
Here we are interested in the cosmological  implications of the model relating to the dark energy problem. In order to satisfy our interest, let us consider the power-law solution in the form 
 \begin{equation}\label{6.1}
 a=a_0t^{\eta},
 \end{equation} 
 where $a_0$ and $\eta$ are contants. Thus,
 \begin{eqnarray}
 \rho&=&3(\mu+\nu)\eta^2t^{-2}+0.5(\mu \alpha a_0^nt^{\eta n}+\nu \beta a_0^mt^{\eta m}),\label{6.2}\\
   p&=& -[(\mu+\nu)(-2\eta+3\eta^2)t^{-2}+\frac{\mu\alpha (n+3)}{6}a_0^{n}t^{\eta n}+\frac{\nu\beta (m+3)}{6}a_0^mt^{\eta m}].\label{6.3}
 \end{eqnarray}
 The EoS parameter reads 
 \begin{equation}\label{6.4}
 \omega= \frac{p}{\rho}=-1-\frac{-2\eta(\mu+\nu)+\frac{\mu\alpha n}{6}a_0^{n}t^{\eta n}+\frac{\nu\beta m}{6}a_0^mt^{\eta m}}{3(\mu+\nu)\eta^2t^{-2}+0.5(\mu \alpha a_0^nt^{\eta n}+\nu \beta a_0^mt^{\eta m})}.
 \end{equation}
 These expressions still content some unknown constant parameters. We assume that these parameters have the following values, namely  $\mu=\nu=1=m=n=\alpha=\beta=a_0$, $\eta=2/3$. TIn this case one has
 \begin{eqnarray}
 \rho&=&\frac{8}{3}t^{-2}+t^{2/3},\label{6.5}\\
   p&=&- \frac{4}{3}t^{2/3},\label{6.6}
 \end{eqnarray}
 so that the EoS takes the form
  \begin{equation}\label{6.7}
 \rho=\frac{512}{81p^3}+\frac{3p}{4}.
 \end{equation}
Furthermore, the EoS parameter becomes 
\begin{equation}\label{6.8}
 \omega(t)= \frac{p}{\rho}=-\frac{4}{3+8t^{-8/3}}=-\frac{4t^{8/3}}{3t^{8/3}+8}.
 \end{equation}
 
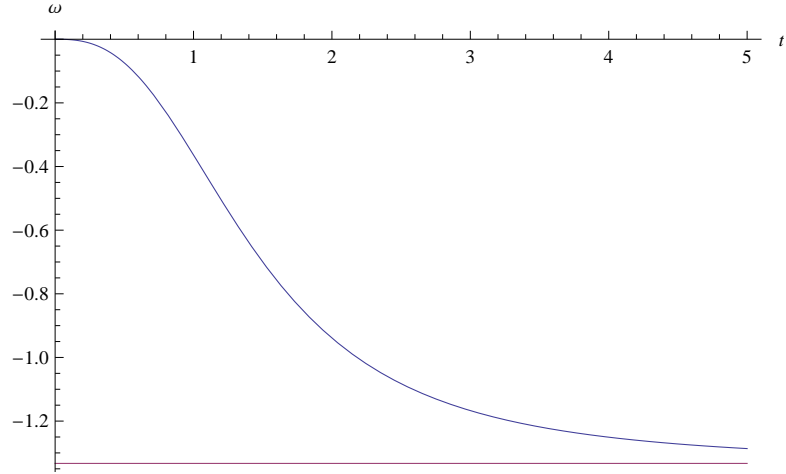
 Hence, we see that  $\omega(0)=0$,  $\omega(1)=-4/11=\approx -0.36$ and $\omega(\infty)=-4/3\approx -1,33$,  so that  our particular case admits the phantom crossing for $\omega=-1$ as $t_0=8^{3/8}$. In Fig.1 we plot the evolution of the EoS parameter with respect to the cosmic time $t$. It is interesting to compare this result with the torsionless case with $\nu=\alpha=\beta=0$, by taking the same values for all the other parameters, namely $\mu=1$ and  $\eta=2/3$, which is the case of GR. As a consequence $p=0$ and $\rho=\frac{8}{3t^{2}}$, which describe the dust matter.

 \section{$F(R,T)$ gravity: Bianchi type I model}
 The results of the section 3 can be extendent to the other metric. As an example, let us consider the M$_{37}$ - model for the Bianchi type spacetime. The corresponding metric is given by
\begin{equation}\label{7.1}
ds^2=-d\tau^2+A^2dx_1^2+B^2dx_2^2+C^2dx_3^2,
\end{equation}
In this case  the M$_{37}$ - model reads as
\begin{eqnarray} \label{7.2}
 S_{39}&=&\int d^4 xe[F(R,T)+L_m],\\
R&=&u+2\left(\frac{\ddot{A}}{A}+\frac{\ddot{B}}{B}+\frac{\ddot{C}}{C}+\frac{\dot{A}\dot{B}}{AB}+\frac{\dot{A}\dot{C}}{AC}+\frac{\dot{B}\dot{C}}{BC}\right),\label{7.3}\\
T&=&v-2\left(\frac{\dot{A}\dot{B}}{AB}+\frac{\dot{A}\dot{C}}{AC}+\frac{\dot{B}\dot{C}}{BC}\right).\label{7.4}
\end{eqnarray} 
  Here $u=u(t, A,B,C,\dot{A},\dot{B}.\dot{C},  \ddot{A},\ddot{B}, \ddot{C},, ...; f_i)$ and $v=v(t, A,B,C,\dot{A},\dot{B}.\dot{C},  \ddot{A},\ddot{B}, \ddot{C},, ...; g_i).$

\section{Other generalizations and reductions  of some generalized  gravity models}

 \subsection{The $F(G)$ with $w$ field}
 Now we consider the  M$_{39}$ - model which looks like 
\begin{eqnarray} 
 S_{39}&=&\int d^4 xe[F(G)+L_m],\label{8.1}\\
G&=&w+24H^2(\dot{H}+H^2),\label{8.2}\\
w&=&w(t, a,\dot{a}, \ddot{a},\dddot{a}, ...; h_i),\label{8.3}
\end{eqnarray} 
 where,   again,   $w=w(t, a,\dot{a}, \ddot{a},\dddot{a}, ...; h_i)$ is  a real function and  $h_i$ is  an unknown function related to the geometry of the spacetime.  If $w=0$ the  M$_{39}$ - model reduces to the usual $F(G)$ gravity with $G=G_s=24H^2(\dot{H}+H^2)$. 
 
  \subsection{The M$_{40}$ - model}
 Now we consider the  M$_{40}$ - model which reads 
 \begin{equation}\label{8.4}
 S_{40}=\int d^4 xe[F(R,G)+L_m],
\end{equation}
where
\begin{eqnarray}
 R&=&u+6(\dot{H}+2H^2),\label{8.5}\\
G&=&w+24H^2(\dot{H}+H^2),\label{8.6}\\
u&=&u(t, a,\dot{a}, \ddot{a},\dddot{a}, ...; f_i),\label{8.7}\\
w&=&w(t, a,\dot{a}, \ddot{a},\dddot{a}, ...; h_i).\label{8.8}
\end{eqnarray}
Here, $u=u(t, a,\dot{a}, \ddot{a},\dddot{a}, ...; f_i)$ and  $w=w(t, a,\dot{a}, \ddot{a},\dddot{a}, ...; h_i)$  are some real functions and  $f_i, h_i, g_i$ are some unknown functions relatedto the geometry of the spacetime.  Note that if we put $u=w=0$, the  M$_{40}$ - model reduces to the usual $F(R,G)$ gravity.

 \subsection{The M$_{38}$ - model}
 Let us consider the following action of the M$_{38}$ - model 
 \begin{equation}\label{8.9}
 {\cal S}_{38}=\int d^4 xe[F(G,T)+L_m],
 \end{equation}
 where \begin{eqnarray}
 G&=&w+24H^2(\dot{H}+H^2),\label{8.10}\\
T&=&v-6H^2,\label{8.11}\\
w&=&w(t, a,\dot{a}, \ddot{a},\dddot{a}, ...; h_i),\label{8.12}\\
v&=&v(t,a,\dot{a}, \ddot{a},\dddot{a}, ...; g_i).\label{8.13}
\end{eqnarray}
Here   in general $w=w(t, a,\dot{a}, \ddot{a},\dddot{a}, ...; h_i)$ and $v=v(t,a,\dot{a}, \ddot{a},\dddot{a}, ...; g_i)$ are some real functions and   $h_i$ and $g_i$ are some unknown functions related with the geometry of the spacetime.

\subsection{The M$_{41}$ - model}
 Now we consider the  M$_{41}$ - model with the following action 
 \begin{equation}\label{8.14}
 {\cal S}_{41}=\int d^4 xe[F(R,G,T)+L_m],
 \end{equation}
 where 
 \begin{eqnarray}
 R&=&u+6(\dot{H}+2H^2),\label{8.15}\\
G&=&w+24H^2(\dot{H}+H^2),\label{8.16}\\
T&=&v-6H^2,\label{8.17}\\
u&=&u(t, a,\dot{a}, \ddot{a},\dddot{a}, ...; f_i),\label{8.18}\\
w&=&w(t, a,\dot{a}, \ddot{a},\dddot{a}, ...; h_i),\label{8.19}\\
v&=&v(t,a,\dot{a}, \ddot{a},\dddot{a}, ...; g_i).\label{8.21}
\end{eqnarray}
Here, again,  $u=u(t, a,\dot{a}, \ddot{a},\dddot{a}, ...; f_i)$, $w=w(t, a,\dot{a}, \ddot{a},\dddot{a}, ...; h_i)$ and $v=v(t,a,\dot{a}, \ddot{a},\dddot{a}, ...; g_i)$ are some real functions and   $f_i, h_i, g_i$ are some unknown functions related to the geometry of the spacetime.

\subsection{The M$_{42}$ - model}
 Let us consider  the M$_{42}$ - model with the action
 \begin{equation}\label{8.21}
 {\cal S}_{42}=\int d^4 xe[F(R,T)+L_m],
 \end{equation}
 where \begin{eqnarray}
 R&=&T\phi+6(\dot{H}+2H^2),\label{8.22}\\
T&=&R\varphi-6H^2\label{8.23}
\end{eqnarray}
Here  $u=T\phi, \quad v=R\varphi$, where  $\phi=\phi(t, a,\dot{a}, \ddot{a},\dddot{a}, ...; \phi_i)$ and $\varphi=\varphi(t, a,\dot{a}, \ddot{a},\dddot{a}, ...; \varphi_i)$ are some unknown functions. 
This model admits (at least) two important particular cases.

\textit{a) The M$_{42A}$ -- model.}
Let us take $R=0$. Then $F(R,T)=F(T)$, \quad $T=-6H^2$ and $\phi=\phi_0=2+H^{-2}\dot{H}$, so that we get purely $F(T)$ gravity.

\textit{b) The M$_{42B}$ -- model.}
Let us take now $T=0$. Then, $F(R,T)=F(R)$, \quad $R=6(\dot{H}+2H^2)$ and $\varphi=\varphi_0=H^2(\dot{H}+2H^2)^{-1}$. This case corresponds to the purely $F(R)$ gravity.

\section{Other examples of metric-affine Myrzakulov gravity theories}
Consider the  metric-affine spacetime with the affine connection $\tilde{\Gamma}^{\lambda}\,_{\mu \nu}$. Then the torsion and nonmetricity tensors are given by
\begin{align}
    T^{\lambda}\,_{\mu \nu}&=2\tilde{\Gamma}^{\lambda}\,_{[\mu \nu]}\,,\\
    Q_{\lambda \mu \nu}&=\tilde{\nabla}_{\lambda}g_{\mu \nu}\,.
\end{align}
The corresponding  covariant derivative of an arbitrary vector $v^{\lambda}$ can be split into a Riemannian contribution and a distortion tensor
\begin{equation}
\tilde{\nabla}_{\mu}v^{\lambda}=\nabla_{\mu}v^{\lambda}+N^{\lambda}\,_{\rho\mu}v^{\rho}\,,
\end{equation}
where 
\begin{equation}
N^{\lambda}\,_{\rho\mu}=K^{\lambda}\,_{\rho\mu}+L^{\lambda}\,_{\rho\mu}.
\end{equation}
Here the contortion and disformation tensors read as
\begin{align}
    K^{\lambda}\,_{\rho\mu}&=\frac{1}{2}(T^{\lambda}\,_{\rho\mu}-T_{\rho}\,^{\lambda}\,_{\mu}-T_{\mu}\,^{\lambda}\,_{\rho})\,,
\end{align}
\begin{align}
    L^{\lambda}\,_{\rho\mu}&=\frac{1}{2}(Q^{\lambda}\,_{\rho\mu}-Q_{\rho}\,^{\lambda}\,_{\mu}-Q_{\mu}\,^{\lambda}\,_{\rho}),
\end{align}
respectively. The commutation of  the covariant derivatives takes the form
\begin{equation}
[\tilde{\nabla}_{\mu},\tilde{\nabla}_{\nu}]\,v^{\lambda}=\tilde{R}^{\lambda}\,_{\rho \mu \nu}\,v^{\rho}+T^{\rho}\,_{\mu \nu}\,\tilde{\nabla}_{\rho}v^{\lambda}\,,
\end{equation}
where
\begin{equation}\label{totalcurvature}
\tilde{R}^{\lambda}\,_{\rho \mu \nu}=\partial_{\mu}\tilde{\Gamma}^{\lambda}\,_{\rho \nu}-\partial_{\nu}\tilde{\Gamma}^{\lambda}\,_{\rho \mu}+\tilde{\Gamma}^{\lambda}\,_{\sigma \mu}\tilde{\Gamma}^{\sigma}\,_{\rho \nu}-\tilde{\Gamma}^{\lambda}\,_{\sigma \nu}\tilde{\Gamma}^{\sigma}\,_{\rho \mu}\,.
\end{equation}
Note that the geometric structure of the metric-affine spacetime is determined by  three tensors: the metric tensor ($g_{\mu\nu}$), the torsion tensor ($T^{\lambda}_{\, \, \, \mu\nu}$) and the nonmetricity tensor ($Q_{\lambda\mu\nu}$). The torsion tensor is the antisymmetric part of the connection and the nonmetricity tensor measures the failure
of the connection to be metric compatible. Note that these three tensors  can be computed
once an affine connection $\tilde{\Gamma}^{\alpha}_{\beta\lambda}$ is given. In this metric-affine  spacetime, let us introduce five  scalars - $R, \, T, \, Q, \, G, \, B$, where $R$ is the metric-affine  curvature scalar, $T$ is the metric-affine torsion scalar,  $Q$ is the metric-affine nonmetricity scalar,    $G$ is the metric-affine Gauss-Bonnet  scalar, $B$ is the boundary term scalar. Below ${\cal T}$ is the trace of the energy-momentum tensor. In the previous sections, we have considered the Myrzakulov gravity-I (MG-I) which has the following action
\begin{eqnarray}
S=\int \sqrt{-g}d^{4}x[F(R,T)+L_{m}],
\end{eqnarray}
where $R$ is the curvature scalar, $T$ is the torsion scalar and $L_{m}$ is the matter Lagrangian. This MG-I is some kind generalization (unification)  of the well-known $F(R)$ and $F(T)$ gravity theories. We now going to present some other examples of   metric-affine Myrzakulov gravity theories, also abbreviated below as MG-N, where N=I, II, III, IV, ... (see, also, Table 1 and Table 2).  

  \subsection{MG-I}
 The action of the Myrzakulov    gravity - I (MG-I) has the following form 
\begin{eqnarray}
S=\frac{1}{2\kappa^{2}}\int \sqrt{-g}d^{4}x[F(R,T)+2\kappa^{2}L_{m}],
\end{eqnarray}
where $R$ is the curvature scalar, $T$ is the torsion scalar and $L_{m}$ is the matter Lagrangian. This MG-I is some kind generalizations of the well-known $F(R)$ and $F(T)$ gravity theories. If exactly, the MG-I is the unification of the $F(R)$ and $F(T)$ theories.    
\subsection{MG-II}
 The action of the Myrzakulov  gravity - II (MG-II) reads as 
\begin{eqnarray}
S=\frac{1}{2\kappa^{2}}\int \sqrt{-g}d^{4}x[F(R,Q)+2\kappa^{2}L_{m}],
\end{eqnarray}
where  $R$ is the curvature scalar and $Q$ is the nonmetricity scalar of the   metric-affine spacetime.
\subsection{MG-III}
 The action of the Myrzakulov   gravity - III (MG-III) reads as 
\begin{eqnarray}
S=\frac{1}{2\kappa^{2}}\int \sqrt{-g}d^{4}x[F(T,Q)+2\kappa^{2}L_{m}],
\end{eqnarray}
where  $T$ is the torsion  scalar and $Q$ is the nonmetricity scalar of the   metric-affine spacetime.
\subsection{MG-IV}
The action of the Myrzakulov  gravity - IV (MG-IV) has the following form
\begin{eqnarray}
S=\frac{1}{2\kappa^{2}}\int \sqrt{-g}d^{4}x[F(R,T,{\cal T})+2\kappa^{2}L_{m}],
\end{eqnarray}
where $R$ is the curvature scalar, $T$ is the torsion scalar and  ${\cal T}$ is the trace of the energy-momentum tensor.

\subsection{MG-V}
The action of the Myrzakulov gravity - V (MG-V) is given by
\begin{eqnarray}
S=\frac{1}{2\kappa^{2}}\int \sqrt{-g}d^{4}x[F(R,T,Q)+2\kappa^{2}L_{m}],
\end{eqnarray}
where $R$ is the curvature scalar, $T$ is the torsion scalar and  $Q$ is the nonmetricity scalar of the   metric-affine spacetime. 
\subsection{MG-VI}
 The action of the Myrzakulov   gravity - VI (MG-VI) reads as 
\begin{eqnarray}
S=\frac{1}{2\kappa^{2}}\int \sqrt{-g}d^{4}x[F(R,Q, {\cal T})+2\kappa^{2}L_{m}],
\end{eqnarray}
where  $R$ is the curvature scalar,  $Q$ is the nonmetricity scalar and  ${\cal T}$ is the trace of the energy-momentum tensor
 of our generalized spacetime.
\subsection{MG-VII}
 The action of the Myrzakulov   gravity - VII (MG-VII) reads as 
\begin{eqnarray}
S=\frac{1}{2\kappa^{2}}\int \sqrt{-g}d^{4}x[F(T,Q, {\cal T})+2\kappa^{2}L_{m}],
\end{eqnarray}
and   $T$ is the torsion  scalar,  $Q$ is the nonmetricity scalar and  ${\cal T}$ is the trace of the energy-momentum tensor
 of the   metric-affine spacetime.

\subsection{MG-VIII}
The action of the Myrzakulov gravity - VIII (MG-VIII) reads as
\begin{eqnarray}
S=\frac{1}{2\kappa^{2}}\int \sqrt{-g}d^{4}x[F(R,T,Q, {\cal T})+2\kappa^{2}L_{m}],
\end{eqnarray}
where $R$ is the curvature  scalar, $T$ is the torsion scalar, $Q$ is the nonmetricity scalar and  ${\cal T}$ is the trace of the energy-momentum tensor (the trace of the stress-energy tensor) of the   metric-affine spacetime.
 
  \subsection{MG-IX}
 The action of the Myrzakulov   gravity - IX (MG-IX) has the following form 
\begin{eqnarray}
S=\frac{1}{2\kappa^{2}}\int \sqrt{-g}d^{4}x[F(R,T,G)+2\kappa^{2}L_{m}],
\end{eqnarray}
where $R$ is the curvature scalar, $T$ is the torsion scalar, $G$ is the metric-affine Gauss-Bonnet scalar  of the   metric-affine spacetime.     
\subsection{MG-X}
 The action of the Myrzakulov gravity - X (MG-X) reads as 
\begin{eqnarray}
S=\frac{1}{2\kappa^{2}}\int \sqrt{-g}d^{4}x[F(R,Q,G)+2\kappa^{2}L_{m}],
\end{eqnarray}
where  $R$ is the curvature scalar,  $Q$ is the nonmetricity scalar, $G$ is the metric-affine Gauss-Bonnet scalar of the metric-affine spacetime.
\subsection{MG-XI}
 The action of the Myrzakulov gravity - XI (MG-XI) reads as 
\begin{eqnarray}
S=\frac{1}{2\kappa^{2}}\int \sqrt{-g}d^{4}x[F(T,Q,G)+2\kappa^{2}L_{m}],
\end{eqnarray}
where  $T$ is the metric-affine torsion  scalar,  $Q$ is the metric-affine  nonmetricity scalar and $G$ is the metric-affine Gauss-Bonnet scalar of our metric-affine  spacetime.
\subsection{MG-XII}
The action of the Myrzakulov  gravity - XII (MG-XII) has the following form
\begin{eqnarray}
S=\frac{1}{2\kappa^{2}}\int \sqrt{-g}d^{4}x[F(R,T,G, {\cal T})+2\kappa^{2}L_{m}],
\end{eqnarray}
where $R$ is the metric-affine curvature scalar, $T$ is the metric-affine torsion scalar, $G$ is the metric-affine Gauss-Bonnet scalar and  ${\cal T}$ is the trace of the energy-momentum tensor.

\subsection{MG-XIII}
The action of the Myrzakulov  gravity - XIII (MG-XIII) is given by
\begin{eqnarray}
S=\frac{1}{2\kappa^{2}}\int \sqrt{-g}d^{4}x[F(R,T,Q,G)+2\kappa^{2}L_{m}],
\end{eqnarray}
where $R$ is the curvature scalar, $T$ is the torsion scalar,   $Q$ is the nonmetricity scalar and $G$ is the metric-affine Gauss-Bonnet scalar of the   metric-affine spacetime.
\subsection{MG-XIV}
 The action of the Myrzakulov   gravity - XIV (MG-XIV) reads as 
\begin{eqnarray}
S=\frac{1}{2\kappa^{2}}\int \sqrt{-g}d^{4}x[F(R,Q, G, {\cal T})+2\kappa^{2}L_{m}],
\end{eqnarray}
where  $R$ is the metric-affine curvature scalar,  $Q$ is the metric-affine nonmetricity scalar, $G$ is the metric-affine Gauss-Bonnet scalar and  ${\cal T}$ is the trace of the energy-momentum tensor
 of the  metric-affine  spacetime.
\subsection{MG-XV}
 The action of the Myrzakulov   gravity - XV (MG-XV) reads as 
\begin{eqnarray}
S=\frac{1}{2\kappa^{2}}\int \sqrt{-g}d^{4}x[F(T,Q,G,  {\cal T})+2\kappa^{2}L_{m}],
\end{eqnarray}
and   $T$ is the metric-affine torsion  scalar,  $Q$ is the metric-affine nonmetricity scalar, $G$ is the metric-affine Gauss-Bonnet scalar and  ${\cal T}$ is the trace of the energy-momentum tensor
 of our metric-affine  spacetime. 

\subsection{MG-XVI}
The action of the Myrzakulov gravity - XVI (MG-XVI) reads as
\begin{eqnarray}
S=\frac{1}{2\kappa^{2}}\int \sqrt{-g}d^{4}x[F(R,T,Q, G, {\cal T})+2\kappa^{2}L_{m}],
\end{eqnarray}
where $R$ is the metric-affine curvature  scalar, $T$ is the metric-affine torsion scalar, $Q$ is the metric-affine nonmetricity scalar, $G$ is the metric-affine Gauss-Bonnet scalar and  ${\cal T}$ is the trace of the energy-momentum tensor  of the  metric-affine  spacetime.
 \subsection{MG-XVII}
The action of the Myrzakulov gravity - XVII (MG-XVII) reads as
\begin{eqnarray}
S=\frac{1}{2\kappa^{2}}\int \sqrt{-g}d^{4}x[F(Q, G)+2\kappa^{2}L_{m}],
\end{eqnarray}
where  $Q$ is the metric-affine nonmetricity scalar and $G$ is the metric-affine Gauss-Bonnet scalar  of the metric-affine  spacetime.

\subsection{MG-XVIII}
The action of the Myrzakulov gravity - XVIII (MG-XVIII) reads as
\begin{eqnarray}
S=\frac{1}{2\kappa^{2}}\int \sqrt{-g}d^{4}x[F(R,T,G)+2\kappa^{2}L_{m}],
\end{eqnarray}
where $R$ is the metric-affine curvature  scalar, $T$ is the metric-affine torsion scalar and  $G$ is the metric-affine Gauss-Bonnet scalar of the metric-affine  spacetime.
\subsection{MG-XIX}
The action of the Myrzakulov gravity - XIX (MG-XIX) reads as
\begin{eqnarray}
S=\frac{1}{2\kappa^{2}}\int \sqrt{-g}d^{4}x[F(T,G, {\cal T})+2\kappa^{2}L_{m}],
\end{eqnarray}
where  $T$ is the metric-affine torsion scalar,  $G$ is the metric-affine Gauss-Bonnet scalar and  ${\cal T}$ is the trace of the energy-momentum tensor  of the  metric-affine  spacetime.

  \subsection{MG-XX}
 The action of the Myrzakulov    gravity - XX (MG-XX) has the following form 
\begin{eqnarray}
S=\frac{1}{2\kappa^{2}}\int \sqrt{-g}d^{4}x[F(R,T,B)+2\kappa^{2}L_{m}],
\end{eqnarray}
where $R$ is the curvature scalar, $T$ is the torsion scalar, $B$ is  the boundary term scalar and $L_{m}$ is the matter Lagrangian. This MG-I is some kind generalizations of the well-known $F(R)$ and $F(T)$ gravity theories. If exactly, the MG-I is the unification of the $F(R)$ and $F(T)$ theories.    
\subsection{MG-XXI}
 The action of the Myrzakulov  gravity - XXI (MG-XXI) reads as 
\begin{eqnarray}
S=\frac{1}{2\kappa^{2}}\int \sqrt{-g}d^{4}x[F(R,Q,B)+2\kappa^{2}L_{m}],
\end{eqnarray}
where  $R$ is the curvature scalar, $B$ is  the boundary term scalar and $Q$ is the nonmetricity scalar of the   metric-affine spacetime.
\subsection{MG-XXII}
 The action of the Myrzakulov   gravity - XXII (MG-XXII) reads as 
\begin{eqnarray}
S=\frac{1}{2\kappa^{2}}\int \sqrt{-g}d^{4}x[F(T,Q, B)+2\kappa^{2}L_{m}],
\end{eqnarray}
where  $T$ is the torsion  scalar, $B$ is  the boundary term scalar and $Q$ is the nonmetricity scalar of the   metric-affine spacetime.
\subsection{MG-XXIII}
The action of the Myrzakulov  gravity - XXIII (MG-XXIII) has the following form
\begin{eqnarray}
S=\frac{1}{2\kappa^{2}}\int \sqrt{-g}d^{4}x[F(R,T,B, {\cal T})+2\kappa^{2}L_{m}],
\end{eqnarray}
where $R$ is the curvature scalar, $T$ is the torsion scalar, $B$ is  the boundary term scalar and  ${\cal T}$ is the trace of the energy-momentum tensor.

\subsection{MG-XXIV}
The action of the Myrzakulov gravity - XXIV (MG-XXIV) is given by
\begin{eqnarray}
S=\frac{1}{2\kappa^{2}}\int \sqrt{-g}d^{4}x[F(R,T,Q, B)+2\kappa^{2}L_{m}],
\end{eqnarray}
where $R$ is the curvature scalar, $T$ is the torsion scalar, $B$ is  the boundary term scalar and  $Q$ is the nonmetricity scalar of the   metric-affine spacetime. 
\subsection{MG-XXV}
 The action of the Myrzakulov   gravity - XXV (MG-XXV) reads as 
\begin{eqnarray}
S=\frac{1}{2\kappa^{2}}\int \sqrt{-g}d^{4}x[F(R,Q, B,  {\cal T})+2\kappa^{2}L_{m}],
\end{eqnarray}
where  $R$ is the curvature scalar,  $Q$ is the nonmetricity scalar, $B$ is  the boundary term scalar and  ${\cal T}$ is the trace of the energy-momentum tensor
 of our generalized spacetime.
\subsection{MG-XXVI}
 The action of the Myrzakulov   gravity - XXVI (MG-XXVI) reads as 
\begin{eqnarray}
S=\frac{1}{2\kappa^{2}}\int \sqrt{-g}d^{4}x[F(T,Q, B, {\cal T})+2\kappa^{2}L_{m}],
\end{eqnarray}
and   $T$ is the torsion  scalar,  $Q$ is the nonmetricity scalar, $B$ is  the boundary term scalar  and  ${\cal T}$ is the trace of the energy-momentum tensor
 of the   metric-affine spacetime.

\subsection{MG-XXVII}
The action of the Myrzakulov gravity - XXVII (MG-XXVII) reads as
\begin{eqnarray}
S=\frac{1}{2\kappa^{2}}\int \sqrt{-g}d^{4}x[F(R,T,Q, B,  {\cal T})+2\kappa^{2}L_{m}],
\end{eqnarray}
where $R$ is the curvature  scalar, $T$ is the torsion scalar, $Q$ is the nonmetricity scalar, $B$ is  the boundary term scalar and  ${\cal T}$ is the trace of the energy-momentum tensor (the trace of the stress-energy tensor) of the   metric-affine spacetime.
 
  \subsection{MG-XXVIII}
 The action of the Myrzakulov   gravity - XXVIII (MG-XXVIII) has the following form 
\begin{eqnarray}
S=\frac{1}{2\kappa^{2}}\int \sqrt{-g}d^{4}x[F(R,T,G, B)+2\kappa^{2}L_{m}],
\end{eqnarray}
where $R$ is the curvature scalar, $T$ is the torsion scalar, $B$ is  the boundary term scalar, $G$ is the metric-affine Gauss-Bonnet scalar  of the   metric-affine spacetime.     
\subsection{MG-XXIX}
 The action of the Myrzakulov gravity - XXIX (MG-XXIX) reads as 
\begin{eqnarray}
S=\frac{1}{2\kappa^{2}}\int \sqrt{-g}d^{4}x[F(R,Q,G, B)+2\kappa^{2}L_{m}],
\end{eqnarray}
where  $R$ is the curvature scalar,  $Q$ is the nonmetricity scalar, $B$ is  the boundary term scalar,  $G$ is the metric-affine Gauss-Bonnet scalar of the metric-affine spacetime.
\subsection{MG-XXX}
 The action of the Myrzakulov gravity - XXX (MG-XXX) reads as 
\begin{eqnarray}
S=\frac{1}{2\kappa^{2}}\int \sqrt{-g}d^{4}x[F(T,Q,G, B)+2\kappa^{2}L_{m}],
\end{eqnarray}
where  $T$ is the metric-affine torsion  scalar,  $Q$ is the metric-affine  nonmetricity scalar, $B$ is  the boundary term scalar and $G$ is the metric-affine Gauss-Bonnet scalar of our metric-affine  spacetime.
\subsection{MG-XXXI}
The action of the Myrzakulov  gravity - XXXI (MG-XXXI) has the following form
\begin{eqnarray}
S=\frac{1}{2\kappa^{2}}\int \sqrt{-g}d^{4}x[F(R,T,G, B, {\cal T})+2\kappa^{2}L_{m}],
\end{eqnarray}
where $R$ is the metric-affine curvature scalar, $T$ is the metric-affine torsion scalar, $G$ is the metric-affine Gauss-Bonnet scalar, $B$ is  the boundary term scalar and  ${\cal T}$ is the trace of the energy-momentum tensor.

\subsection{MG-XXXII}
The action of the Myrzakulov  gravity - XXXII (MG-XXXII) is given by
\begin{eqnarray}
S=\frac{1}{2\kappa^{2}}\int \sqrt{-g}d^{4}x[F(R,T,Q,G, B)+2\kappa^{2}L_{m}],
\end{eqnarray}
where $R$ is the curvature scalar, $T$ is the torsion scalar,   $Q$ is the nonmetricity scalar, $B$ is  the boundary term scalar and $G$ is the metric-affine Gauss-Bonnet scalar of the   metric-affine spacetime.
\subsection{MG-XXXIII}
 The action of the Myrzakulov   gravity - XXXIII (MG-XXXIII) reads as 
\begin{eqnarray}
S=\frac{1}{2\kappa^{2}}\int \sqrt{-g}d^{4}x[F(R,Q, G, B,  {\cal T})+2\kappa^{2}L_{m}],
\end{eqnarray}
where  $R$ is the metric-affine curvature scalar,  $Q$ is the metric-affine nonmetricity scalar, $G$ is the metric-affine Gauss-Bonnet scalar, $B$ is  the boundary term scalar and  ${\cal T}$ is the trace of the energy-momentum tensor
 of the  metric-affine  spacetime.
\subsection{MG-XXXIV}
 The action of the Myrzakulov   gravity - XXXIV (MG-XXXIV) reads as 
\begin{eqnarray}
S=\frac{1}{2\kappa^{2}}\int \sqrt{-g}d^{4}x[F(T,Q,G, B,  {\cal T})+2\kappa^{2}L_{m}],
\end{eqnarray}
and   $T$ is the metric-affine torsion  scalar,  $Q$ is the metric-affine nonmetricity scalar, $G$ is the metric-affine Gauss-Bonnet scalar, $B$ is  the boundary term scalar and  ${\cal T}$ is the trace of the energy-momentum tensor
 of our metric-affine  spacetime. 

\subsection{MG-XXXV}
The action of the Myrzakulov gravity - XXXV (MG-XXXV) reads as
\begin{eqnarray}
S=\frac{1}{2\kappa^{2}}\int \sqrt{-g}d^{4}x[F(R,T,Q, G, B,  {\cal T})+2\kappa^{2}L_{m}],
\end{eqnarray}
where $R$ is the metric-affine curvature  scalar, $T$ is the metric-affine torsion scalar, $Q$ is the metric-affine nonmetricity scalar, $G$ is the metric-affine Gauss-Bonnet scalar, $B$ is  the boundary term scalar and  ${\cal T}$ is the trace of the energy-momentum tensor  of the  metric-affine  spacetime.
 \subsection{MG-XXXVI}
The action of the Myrzakulov gravity - XXXVI (MG-XXXVI) reads as
\begin{eqnarray}
S=\frac{1}{2\kappa^{2}}\int \sqrt{-g}d^{4}x[F(Q, G, B)+2\kappa^{2}L_{m}],
\end{eqnarray}
where  $Q$ is the metric-affine nonmetricity scalar, $B$ is  the boundary term scalar and $G$ is the metric-affine Gauss-Bonnet scalar  of the metric-affine  spacetime.

\subsection{MG-XXXVII}
The action of the Myrzakulov gravity - XXXVII (MG-XXXVII) reads as
\begin{eqnarray}
S=\frac{1}{2\kappa^{2}}\int \sqrt{-g}d^{4}x[F(R,T,G, B)+2\kappa^{2}L_{m}],
\end{eqnarray}
where $R$ is the metric-affine curvature  scalar, $T$ is the metric-affine torsion scalar, $B$ is  the boundary term scalar and  $G$ is the metric-affine Gauss-Bonnet scalar of the metric-affine  spacetime.
\subsection{MG-XXXVIII}
The action of the Myrzakulov gravity - XXXVIII (MG-XXXVIII) reads as
\begin{eqnarray}
S=\frac{1}{2\kappa^{2}}\int \sqrt{-g}d^{4}x[F(T,G, B, {\cal T})+2\kappa^{2}L_{m}],
\end{eqnarray}
where  $T$ is the metric-affine torsion scalar,  $G$ is the metric-affine Gauss-Bonnet scalar, $B$ is  the boundary term scalar and  ${\cal T}$ is the trace of the energy-momentum tensor  of the  metric-affine  spacetime.

\newpage

\begin{table}
\caption{Metric-affine Myrzakulov gravity theories}
\begin{tabular}{|c|c|c|}
\hline 
N & Name & Action \\ 
\hline 
1 &Myrzakulov Gravity - I (MG-I) & $S=\frac{1}{2k^2}\int d^4x \sqrt{-g}\left[F(R,T)+2k^{2}L_m\right]$ \\ 
\hline 
2 &Myrzakulov Gravity - II (MG-II) & $S=\frac{1}{2k^2}\int d^4x \sqrt{-g}\left[F(R,Q)+2k^{2}L_m\right]$\\ 
\hline 
3 &Myrzakulov Gravity - III (MG-III) & $S=\frac{1}{2k^2}\int d^4x \sqrt{-g}\left[F(T,Q)+2k^{2}L_m\right]$ \\ 
\hline 
4 & Myrzakulov Gravity - IV (MG-IV) & $S=\frac{1}{2k^2}\int d^4x \sqrt{-g}\left[F(R,T,{\cal T})+2k^{2}L_m\right]$ \\ 
\hline 
5 & Myrzakulov Gravity - V (MG-V) & $S=\frac{1}{2k^2}\int d^4x \sqrt{-g}\left[F(R,T,Q)+2k^{2}L_m\right]$ \\ 
\hline 
6 & Myrzakulov Gravity - VI (MG-VI) & $S=\frac{1}{2k^2}\int d^4x \sqrt{-g}\left[F(R,Q,{\cal T})+2k^{2}L_m\right]$ \\ 
\hline 
7 & Myrzakulov Gravity - VII (MG-VII) & $S=\frac{1}{2k^2}\int d^4x \sqrt{-g}\left[F(T,Q,{\cal T})+2k^{2}L_m\right]$ \\ 
\hline 
8 & Myrzakulov Gravity - VIII (MG-VIII) & $S=\frac{1}{2k^2}\int d^4x \sqrt{-g}\left[F(R,T,Q,{\cal T})+2k^{2}L_m\right]$ \\ 
\hline  
9 & Myrzakulov Gravity - IX (MG-IX) & $S=\frac{1}{2k^2}\int d^4x \sqrt{-g}\left[F(R,T,G)+2k^{2}L_m\right]$ \\ 
\hline 
10 &Myrzakulov Gravity - X (MG-X) & $S=\frac{1}{2k^2}\int d^4x \sqrt{-g}\left[F(R,Q,G)+2k^{2}L_m\right]$\\ 
\hline 
11 &Myrzakulov Gravity - XI (MG-XI) & $S=\frac{1}{2k^2}\int d^4x \sqrt{-g}\left[F(T,Q,G)+2k^{2}L_m\right]$ \\ 
\hline 
12 & Myrzakulov Gravity - XII (MG-XII) & $S=\frac{1}{2k^2}\int d^4x \sqrt{-g}\left[F(R,T,G,{\cal T})+2k^{2}L_m\right]$ \\ 
\hline 
13 & Myrzakulov Gravity - XIII (MG-XIII) & $S=\frac{1}{2k^2}\int d^4x \sqrt{-g}\left[F(R,T,Q,G)+2k^{2}L_m\right]$ \\ 
\hline 
14 & Myrzakulov Gravity - XIV (MG-XIV) & $S=\frac{1}{2k^2}\int d^4x \sqrt{-g}\left[F(R,Q,G,{\cal T})+2k^{2}L_m\right]$ \\ 
\hline 
15 & Myrzakulov Gravity - XV (MG-XV) & $S=\frac{1}{2k^2}\int d^4x \sqrt{-g}\left[F(T,Q,G,{\cal T})+2k^{2}L_m\right]$ \\ 
\hline 
16 & Myrzakulov Gravity - XVI (MG-XVI) & $S=\frac{1}{2k^2}\int d^4x \sqrt{-g}\left[F(R,T,Q,G,{\cal T})+2k^{2}L_m\right]$ \\ 
\hline  
17 & Myrzakulov Gravity - XVII (MG-XVII) & $S=\frac{1}{2k^2}\int d^4x \sqrt{-g}\left[F(Q,G)+2k^{2}L_m\right]$ \\ 
\hline 
18 & Myrzakulov Gravity - XVIII (MG-XVIII) & $S=\frac{1}{2k^2}\int d^4x \sqrt{-g}\left[F(R,T,G)+2k^{2}L_m\right]$ \\ 
\hline  
19 & Myrzakulov Gravity - XIX (MG-XIX) & {$S=\frac{1}{2k^2}\int d^4x \sqrt{-g}\left[F(T,G,{\cal T})+2k^{2}L_m\right]$} \\ 
\hline 
\end{tabular} 
\end{table}

\newpage

\begin{table}
\caption{Metric-affine MG theories  with the boundary term scalars}
\begin{tabular}{|c|c|c|}
\hline 
N & Name & Action \\ 
\hline 
1 &Myrzakulov Gravity - XX (MG-XX) & $S=\frac{1}{2k^2}\int d^4x \sqrt{-g}\left[F(R,T,B)+2k^{2}L_m\right]$ \\ 
\hline 
2 &Myrzakulov Gravity - XXI (MG-XXI) & $S=\frac{1}{2k^2}\int d^4x \sqrt{-g}\left[F(R,Q,B)+2k^{2}L_m\right]$\\ 
\hline 
3 &Myrzakulov Gravity - XXII (MG-XXII) & $S=\frac{1}{2k^2}\int d^4x \sqrt{-g}\left[F(T,Q,B)+2k^{2}L_m\right]$ \\ 
\hline 
4 & Myrzakulov Gravity - XXIII (MG-XXIII) & $S=\frac{1}{2k^2}\int d^4x \sqrt{-g}\left[F(R,T,B,{\cal T})+2k^{2}L_m\right]$ \\ 
\hline 
5 & Myrzakulov Gravity - XXIV (MG-XXIV) & $S=\frac{1}{2k^2}\int d^4x \sqrt{-g}\left[F(R,T,Q,B)+2k^{2}L_m\right]$ \\ 
\hline 
6 & Myrzakulov Gravity - XXV (MG-XXV) & $S=\frac{1}{2k^2}\int d^4x \sqrt{-g}\left[F(R,Q,B,{\cal T})+2k^{2}L_m\right]$ \\ 
\hline 
7 & Myrzakulov Gravity - XXVI (MG-XXVI) & $S=\frac{1}{2k^2}\int d^4x \sqrt{-g}\left[F(T,Q,B, {\cal T})+2k^{2}L_m\right]$ \\ 
\hline 
8 & Myrzakulov Gravity - XXVII (MG-XXVII) & $S=\frac{1}{2k^2}\int d^4x \sqrt{-g}\left[F(R,T,Q,B,{\cal T})+2k^{2}L_m\right]$ \\ 
\hline  
9 & Myrzakulov Gravity - XXVIII (MG-XXVIII) & $S=\frac{1}{2k^2}\int d^4x \sqrt{-g}\left[F(R,T,G,B)+2k^{2}L_m\right]$ \\ 
\hline 
10 &Myrzakulov Gravity - XXIX (MG-XXIX) & $S=\frac{1}{2k^2}\int d^4x \sqrt{-g}\left[F(R,Q,G,B)+2k^{2}L_m\right]$\\ 
\hline 
11 &Myrzakulov Gravity - XXX (MG-XXX) & $S=\frac{1}{2k^2}\int d^4x \sqrt{-g}\left[F(T,Q,G,B)+2k^{2}L_m\right]$ \\ 
\hline 
12 & Myrzakulov Gravity - XXXI (MG-XXXI) & $S=\frac{1}{2k^2}\int d^4x \sqrt{-g}\left[F(R,T,G,B,{\cal T})+2k^{2}L_m\right]$ \\ 
\hline 
13 & Myrzakulov Gravity - XXXII (MG-XXXII) & $S=\frac{1}{2k^2}\int d^4x \sqrt{-g}\left[F(R,T,Q,G,B)+2k^{2}L_m\right]$ \\ 
\hline 
14 & Myrzakulov Gravity - XXXIII (MG-XXXIII) & $S=\frac{1}{2k^2}\int d^4x \sqrt{-g}\left[F(R,Q,G,B,{\cal T})+2k^{2}L_m\right]$ \\ 
\hline 
15 & Myrzakulov Gravity - XXXIV (MG-XXXXIV) & $S=\frac{1}{2k^2}\int d^4x \sqrt{-g}\left[F(T,Q,G,B,{\cal T})+2k^{2}L_m\right]$ \\ 
\hline 
16 & Myrzakulov Gravity - XXXV (MG-XXXV) & $S=\frac{1}{2k^2}\int d^4x \sqrt{-g}\left[F(R,T,Q,G,B,{\cal T})+2k^{2}L_m\right]$ \\ 
\hline  
17 & Myrzakulov Gravity - XXXVI (MG-XXXVI) & $S=\frac{1}{2k^2}\int d^4x \sqrt{-g}\left[F(Q,G,B)+2k^{2}L_m\right]$ \\ 
\hline 
18 & Myrzakulov Gravity - XXXVII (MG-XXXVII) & $S=\frac{1}{2k^2}\int d^4x \sqrt{-g}\left[F(R,T,G,B)+2k^{2}L_m\right]$ \\ 
\hline  
19 & Myrzakulov Gravity - XXXVIII (MG-XXXVIII) & {$S=\frac{1}{2k^2}\int d^4x \sqrt{-g}\left[F(T,G,B,{\cal T})+2k^{2}L_m\right]$} \\ 
\hline 
\end{tabular} 
\end{table}

\section{Cosmology in MG theories}
Consider the FRW universe.  The flat FRW spacetime is described by the metric
 \begin{equation}\label{1.1}
 ds^2=-dt^2+a^2(t)(dx^2+dy^2+dz^2),
 \end{equation}
 where $a=a(t)$ is the scale factor.
Let $R$, $T$, $Q$ are the Ricci, torsion,  nonmetricity  scalars. For the FRW metric they have the forms: i)  $R=R_{0}$, where $T=Q=0$;  ii) $T=T_{0}$, where $R=Q=0$; iii) $Q=Q_{0}$, where $R=T=0$. For the FRW metric, they have the forms: 
\begin{eqnarray}
 R_{0}&=&6(\dot{H}+2H^2),\\
T_{0}&=&-6H^2,\\
 Q_{0}&=& 6H^2,
   \end{eqnarray}  
 where $H=(\ln a)_{t}$  is the Hubble parameter. In the  metric-affine spacetime case, we assume that the Ricci, torsion and nonmetricity scalars take  the forms 
   \begin{eqnarray}
 R&=&6(\dot{H}+2H^2)+u,\\
   T&=&-6H^2+v,\\
	Q&=&6H^{2}+w.
 \end{eqnarray}
 Similarly, we can write the boundary term scalar ($B$) and the GB scalar as
\begin{eqnarray}
 G&=&G_{0}+p,\\
   B&=&B_{0}+f,
 \end{eqnarray}
where $u, v, w, p, f$ are some real functions of $t, a, \dot{a}, \ddot{a}$. 
\section{Spherically symmetric and black hole solutions in MG theories}
Let us we now  present our idea to study,  for example, the black hole solutions of MG theories. For this aim, we consider the following static and spherically symmetric metric
\begin{eqnarray}
ds^{2}= A(r)dt^{2}-B(r)dr^{2}-C(r) (d\theta^{2}+\sin^{2}\theta d\phi^{2}),
\end{eqnarray}
where $A(r)$, $B(r)$  and $C(r)$ are real functions of the radial coordinate $r$. Consider two connections: the Levi-Civita connection and the Weitzenb{\"{o}}ck connection.
First, let us consider the  Levi-Civita connection case. In this case, the nonmetricity  and torsion scalars are  equal to zero that is $T_{0}=Q_{0}=0$. Then   the corresponding Ricci scalar has the form
\begin{eqnarray}
R_{0}=\frac{A^{\prime\prime}}{AB}+2\frac{C^{\prime\prime}}{BC}+\frac{A^{\prime}C^{\prime}}{ABC}-\frac{A^{\prime 2}}{2A^{2}B}-\frac{C^{\prime 2}}{2BC^{2}}-
\frac{A^{\prime}B^{\prime}}{2AB^{2}}
-\frac{B^{\prime}C^{\prime}}{B^{2}C}-\frac{2}{C}.
\end{eqnarray}
Here and below primes denote differentiation with respect to the radial coordinate $r$. Let us we now  consider  the Weitzenb{\"{o}}ck connection case. In this case,  the Ricci scalar and nonmetricity scalar are  equal to zero that is $R_{0}=Q_{0}=0$ and  the
 torsion scalar is given by
\begin{eqnarray}
T_{0}=\frac{C^{\prime}(2A^{\prime}C+AC^{\prime}}{2ABC^{2}}.
\end{eqnarray}
Similarly, we can calculate  the nonmetricity scalar $Q_{0}$. 		For the metric (11.1), it has the form
\begin{eqnarray}
Q_{0}=-\frac{C^{\prime}(2A^{\prime}C+AC^{\prime}}{2ABC^{2}},
\end{eqnarray}
where $R_{0}=T_{0}=0$. 
The geometry of the  MG theories is the metric-affine spacetime. For that reason, now let us consider the more general case, namely, the metric-affine spacetime. For this metric-affine spacetime, we have the  metric-affine connection. In this metric-affine case, the Ricci scalar,
the torsion scalar
and the nonmetricity scalar take the forms
\begin{eqnarray}
R&=&R_{0}+u, \\
T&=&T_{0}+v, \\
Q&=&Q_{0}+w.
\end{eqnarray}
Here the metric-affine contributions are given by the following functions
\begin{eqnarray}
u&=&u(A,B,C,A^{\prime},B^{\prime},C^{\prime},A^{\prime\prime},B^{\prime\prime},C^{\prime\prime}), \\
v&=&v(A,B,C,A^{\prime},B^{\prime},C^{\prime},A^{\prime\prime},B^{\prime\prime},C^{\prime\prime}), \\
w&=&w(A,B,C,A^{\prime},B^{\prime},C^{\prime},A^{\prime\prime},B^{\prime\prime},C^{\prime\prime}).
\end{eqnarray}
They are some real functions of the metric tensor components $g_{ij}$  (11.1). 
\section{MG theories  with the boundary term scalars}
Next, we very briefly mention the main moments of MG theories  with the boundary term scalars. According our idea, we assume that  the boundary term scalar has the form
\begin{eqnarray}
B=B_{0}+f.
\end{eqnarray}
Similarly, we can write the GB scalar for the metric-affine spacetime as
\begin{eqnarray}
G=G_{0}+p.
\end{eqnarray}
In the last two equations, $p$ and $f$ are metric-affine contributions and some functions of $A,B,C$ and their derivatives. 
 \section{Conclusion}

As it is well known, modified gravity theories  play an important role in modern cosmology. In particular, the well-known $F(R)$ and $F(T)$ theories are useful tools to study dark energy phenomena  motivated at a
 fundamental level. In the present work, we have considered the more general theory, namely the $F(R,T)$- models. 

At first, we have written the equations of the model and we have found their several reductions.  In particular, the
 Lagrangian has been explicitly constructed. The
 corresponding exact solutions are found for the specific model  $F(R,T)=\mu R+\nu T$ theory, for which the
 universe expands as
 $a(t)=a_0 t^{\eta}$. Furthermore, we have considered the physical
 quantities corresponding to the exact solution, and we have found that
 it can describe  the expansion of our universe in an accelerated way
 without introducing the dark energy. 

Some remarks are in order. 
  Of course many aspects of $F(R,T)$ theory are actually unexplored. For example,  we do not have any realistic
 model which fits the cosmological data, unlike $F(R)$ or $F(T)$ theory. 
We do not know viability conditions of the models, 
 , what forms of
 $F(R,T)$ can be derived from fundamental theories and
 so on (it may be extremely 
 important to reconstract a $F(R,T)$-theory by starting from some basical principles). 
On the other hand, we have here 
 shown that the $F(R,T)$ models can be serious candidates as modified gravity models for the dark energy. Also we note that  the behaviour and the results of  $F(R,T)$-gravity may be extremely different with respect to the ones of 
GR, $F(R)$ and $F(T)$ gravity theories,
so that only the observation of our universe 
may discriminate
between the various gravity theories. 
We not want here discuss merits and demerits of the models above, since we think that it requires some more accurate investigations related to cosmological applications.

 \end{document}